\documentclass{article}

\usepackage{arxiv}

\usepackage[utf8]{inputenc} % allow utf-8 input
\usepackage[T1]{fontenc}    % use 8-bit T1 fonts
\usepackage{hyperref}       % hyperlinks
\usepackage{url}            % simple URL typesetting
\usepackage{booktabs}       % professional-quality tables
\usepackage{amsfonts}       % blackboard math symbols
\usepackage{nicefrac}       % compact symbols for 1/2, etc.
\usepackage{microtype}      % microtypography
\usepackage{lipsum}		% Can be removed after putting your text content
\usepackage{graphicx}
\usepackage{natbib}
\usepackage{doi}

\usepackage{enumitem}
\usepackage{amsmath, amsthm}
\usepackage{subcaption}
\usepackage[table]{xcolor}
\usepackage{multirow}
\usepackage{adjustbox}

\title{Event-driven type design for clinical trials with recurrent events}

\date{February 7, 2026}	% Here you can change the date presented in the paper title
\author{ \href{https://orcid.org/0000-0003-2327-8278}{\includegraphics[scale=0.06]{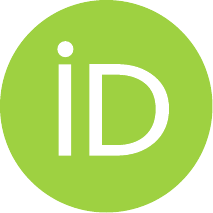}\hspace{1mm}Jingwen Zhang}\\
	Department of Biomedical Statistics, Graduate School of Medicine\\
	The University of Osaka\\
	Yamadaoka 2-2, Suita City, Osaka 565-0871, Japan \\
	\texttt{zhang\_jingwen@biostat.med.osaka-u.ac.jp} \\
	\And
	\href{https://orcid.org/0000-0002-0200-1408}{\includegraphics[scale=0.06]{orcid.pdf}\hspace{1mm}Satoshi Hattori} \\
	Department of Biomedical Statistics, Graduate School of Medicine and\\
    Integrated Frontier Research for Medical Science Division, \\ Institute for Open and Transdisciplinary Research Initiatives (OTRI)\\
	The University of Osaka\\
	Yamadaoka 2-2, Suita City, Osaka 565-0871, Japan\\
	\texttt{hattoris@biostat.med.osaka-u.ac.jp} \\
}

\renewcommand{\shorttitle}{Event-driven type design for clinical trials with recurrent events}

\hypersetup{
pdftitle={Event-driven type design for clinical trials with recurrent events},
pdfsubject={q-bio.NC, q-bio.QM},
pdfauthor={Jingwen Zhang, Satoshi Hattori},
pdfkeywords={Continuous monitoring, Event-driven trial design, Recurrent events, Sandwich variance, Within-subject correlation},
}

\begin{document}
\maketitle

\begin{abstract}
	It is a common practice in randomized clinical trials with the standard survival outcome to follow patients until a prespecified number of events have been observed, a type of trial known as the event-driven trial. The event-driven design ensures that the target power for a specified type 1 error rate is achieved to detect the target hazard ratio, regardless of the specification of other quantities. To understand the treatment effect for chronic conditions, the analysis of recurrent events has gained popularity in randomized controlled trials, particularly large-scale confirmatory trials. In the absence of within-subject correlation among multiple events, a similar event-driven design can be employed for recurrent event outcomes. On the other hand, in the presence of the within-subject correlation, one needs to model the correlation among recurrent events in evaluating power and setting the sample size. However, information useful in modeling the within-subject correlation is limited at the design stage. Failing to consider the correlation properly may lead to underpowered studies. We propose an event-driven type design for recurrent event outcomes. Our method ensures the target power for the target treatment effect, regardless of the specification of other quantities, by monitoring the robust variance under the marginal rates/means model in a blinded manner. We investigate the operating characteristics of the proposed monitoring procedure in simulation studies. The results of simulation studies showed that the proposed blinded monitoring procedure controlled the power well so that the test possessed the target power and did not lead to serious inflation of the type 1 error rate. Furthermore, we illustrate the proposed method using a real clinical trial dataset.
\end{abstract}

% keywords can be removed

\section{Introduction}\label{sec1}
In randomized clinical trials with the standard univariate time-to-event endpoint, the logrank test with the hazard ratio (HR) as the treatment contrast is widely and routinely used. The wide acceptance of the logrank-HR strategy is mainly due to the success of the clever and stable partial likelihood inference for the Cox proportional hazards model. Under the logrank-HR strategy, the event-driven design can be employed. Based on Schoenfeld's power formula of logrank test~\citep{Schoenfeld1981,Schoenfeld1983}, one can derive the target number of events for a given treatment effect at the design stage. A notable feature of the event-driven design is that the target power for the target treatment effect is ensured, regardless of the specification of nuisance quantities, provided the analysis is conducted when the prespecified number of events is observed. Furthermore, since the time of the statistical analysis can be determined without knowledge of the allocation of treatment to subjects, an event-driven study can be implemented with blinded treatment codes. Thus, the event-driven design should contribute to the broad applicability of the logrank-HR strategy and minimize the risk of operational biases.

The advent of advanced therapies and innovative drugs has reshaped the clinical landscape. Many diseases that used to be fatal have now evolved into manageable chronic conditions featured by multiple nonfatal events. This paradigm shift has motivated many researchers to think beyond the univariate time-to-event endpoint, choosing recurrent events as the primary endpoint for randomized controlled trials~\citep{Mogensen2018, Ponikowski2020, Bhatt2021}. Recurrent events are defined as events that can occur repeatedly for the same patient over time. Compared with the standard survival analysis, the statistical properties of recurrent event analysis include the potential to capture the disease burden more effectively, increase statistical power, and reduce sample size~\citep{Gregson2023, Claggett2018, Furberg2022}.

Andersen and Gill generalized the Cox proportional hazards model to recurrent events. The Andersen-Gill (AG) model is intensity-based, modeling the instantaneous incidence of a new event at a specific time \textit{t} given the individual's entire event history up to \textit{t}. The AG model assumes that the time increments between events for a patient are independent, corresponding to a Poisson process. Here, we name the assumption the Poisson-type assumption in the rest of this paper. With the Poisson-type assumption, ~\citet{PatriciaBernardo2001} derived the expected number of events under a homogeneous Poisson process. Using their formula, an event-driven design can be conducted as in the standard survival analysis~\citep{PatriciaBernardo2001}. The Poisson-type assumption imposes a specific dependence structure of recurrent events. However, the within-subject dependence of event times can be more complicated. The risk of ignoring such within-subject dependence includes the inflation of Type 1 error~\citep {Amorim2015}. Thus, for the analysis and design of clinical trials with recurrent events, the within-subject dependence among events should be handled appropriately. \citet{Lin2000} developed the robust sandwich variance, which was valid in the presence of within-subject dependence. 

On the other hand, however, almost all existing sample size calculation methods accounting for within-subject correlation gave formulas for the number of subjects rather than events under some strong assumptions of the probabilistic structure of the recurrent events. If the focus of trials is only the event counts over a prespecified period, sample size formulas proposed by~\citet{TANGO2009466} and~\citet{Li02012015} can be considered. Both discussed the violation of the Poisson-type assumption because data from recurrent events often exhibit overdispersion relative to a Poisson process. \citet{TANGO2009466} recommended the use of robust sandwich variance from a conditional Poisson model, while~\citet{Li02012015} assumed that the variance is proportional to the mean of event counts. Modeling events via a homogeneous Poisson process, Cook (1995) proposed parametric calculations for the expected number of subjects or trial duration under the Poisson regression model. Additionally, ~\citet{Cook1995} modified his formula by introducing subject-specific random effects into the modeling of event counts. \citet{Matsui2005} developed asymptotic sample size formulae for a nonparametric test comparing the rate functions of two groups under a mixed Poisson process (MPP) with a Weibull baseline intensity, explicitly incorporating frailty to account for within-subject dependence. \citet{Tang2019} provided power and sample size formulas for robust Wald tests from the AG model assuming a negative binomial distribution for event counts, covering superiority, noninferiority, and equivalence settings. 

All the methods discussed in the last paragraph set the target number of subjects. As an exceptional event-driven study for recurrent events, \citet{Ingel2014} extended Bernardo and Harrington (2001)'s method to overdispersed data by adopting a frailty‑based inflation factor derived from the decomposition of the robust variance under the AG model proposed by ~\citet{Al-Khalidi2011}. However, the target number of events derived by Ingel and \citet{Ingel2014} depended on specific assumptions for recurrent event processes at the design stage, which are vulnerable to misspecification of the level of dispersion, baseline rate functions, follow-up schemes, etc. Even if the target treatment effect is fixed, it may not ensure the target power since there is limited prior knowledge on the probabilistic structure of the recurrent events. 

To address this particular issue, several strategies for sample size re-estimation have been proposed to mitigate parameter misspecification at the design stage. \citet{cook2009} proposed a two-stage design to revise the sample size with complete or partial blinding, in which an EM algorithm was used to re-estimate the constant baseline intensity function and the frailty variance at an interim analysis. Likewise, ~\citet{Ingel2014} proposed a blinded sample size re-calculation method to update the specification of the frailty variance. However, their methods still rely on the MPP. Besides, the reestimation of the frailty variance only uses data from an internal pilot study, which is a small subset of the recurrent events within a limited time window. To make full use of the accumulating data and get rid of assumptions regarding event rate forms and within-subject dependence, we propose a blind continuous monitoring procedure built on the semiparametric marginal means/rates model. The monitoring threshold is computed using only the anticipated treatment effect, which mirrors the simplicity of classic event‑driven designs. The robust variance proposed by ~\citet{Lin2000} under the semiparametric model is used to drive the stopping criteria and is estimated nonparametrically from pooled data.          

The remainder of this article is organized as follows. We first present the notation and statistical models for analyzing recurrent events, and then provide the power formula and details of the blinded monitoring procedure in Section~\ref{sec1} - ~\ref{sec3}. In Section~\ref{sec4}, we assess the proposed method's operating characteristics and compare it with a conventional event-driven design in simulation studies. Section~\ref{sec5} illustrates the practical implementation using a dataset from a chronic disease trial. Finally, we conclude with a brief discussion.

\section{Preliminary}\label{sec2}

\subsection{Setting and Notation}\label{sub21}

We consider a two-arm clinical trial in which the primary outcome is a recurrent event. A total of $n$ patients are randomized to an experimental or a control group. The allocation of treatment of the subject $i$ is defined by a binary random variable $Z$ with $P(Z=1) = \pi$, which takes the value 1 for the experimental group and 0 for the control group. Each subject can experience the event of interest multiple times. Denote $N^*(t)$ the underlying counting process for events of interest, which is defined by the number of events observed over the interval $[0, t]$. Events are subject to right-censoring by the censoring time $C$, and $N(t)=N^*(t\wedge C)$ is the observed number of events. $Y(t)=I(C\geq t)$ denotes the at-risk process. Hence, the observed data consist of  $n$ independently and identically distributed copies of $\{N(\cdot), Y(\cdot), Z \}$. They are denoted by $\{N_i(\cdot), Y_i(\cdot), Z_i \}$ for $i=1,...,n$, where the subscript of $i$ represents the quantity for the $i$th subject. Suppose that subjects enter the trial within an accrual period of duration $\tau_{a}$ years and are followed for at least $\tau_{f}$ years, resulting in a total study duration of $\tau = \tau_{a} + \tau_{f}$ years. We assume that $P(\tau\geq C)>0$.

\subsection{Andersen-Gill Model and Marginal Rates Model}\label{sub22}

We begin with the Andersen–Gill (AG) model~\citep{Andersen1982}, though our primary focus is the marginal rates model. It is an extension of the Cox proportional hazards model to recurrent event data. Although the AG model allows us to incorporate multiple explanatory variables, we focus on a comparison of two groups, including a single covariate $Z$. The intensity function $\lambda_{Z}(t)$ is defined by
\begin{eqnarray*}
E\{dN^*(t)|\mathcal{F}_t\}=\lambda_{Z}(t)dt,
\end{eqnarray*}
where $\mathcal{F}_t$ is a filtration generated by $\{N^*(s), Z: 0 \le s\le t\}$. The AG intensity model is defined by
\begin{eqnarray}
\lambda_Z(t)=\lambda_0(t) \exp{(\beta Z)},
\label{cox0}
\end{eqnarray}
where $\lambda_0(t)$ is an unspecified continuous baseline intensity function and $\beta$ is an unknown regression coefficient representing a treatment effect, whose true value is denoted by $\beta_0$. This formulation implies that $N^*(t)$ is a time-transformed Poisson process and possesses the independent increment structure~\citep{Lin2000}. One can estimate the regression coefficient by maximizing the partial likelihood function~\citep{Cox1972}. It is obtained by solving the score equation $U(\beta)=0$, where 
\begin{equation}
\label{pscore}
    U(\beta)=\sum_{i=1}^n\int_0^{\tau}(Z_i-\bar{Z}(\beta,u))dN_i(u),
\end{equation}
$S^{(r)}(\beta,t)=n^{-1}\sum_{i=1}^n Y_i(t)Z_i^r exp(\beta Z_i)$, $0 \leq t \leq \tau$ and $\bar{Z}(\beta,t)=\tfrac{S^{(1)}(\beta,t)}{S^{(0)}(\beta,t)}$. The resulting estimator is denoted by $\hat{\beta}$. Andersen and Gill (1982) showed that $n^{1/2}(\hat{\beta}-\beta_0)$ is asymptotically zero mean normal with covariance $A^{-1}$,
\begin{equation}
\label{fisherinf0}
    A = E\Bigl[\int_0^\tau (Z-\bar z(\beta_0,u))^2Y(u)exp(\beta_0Z)d\Lambda_0(t)\Bigl],
\end{equation}
where $\Lambda_0(t)=\int_0^t \lambda_0(u)du$ is the cumulative baseline intensity function and $\bar{z}(\beta,t)$ the stochastic limit of $\bar{Z}(\beta,t)$. 

As mentioned, the standard AG model requires the independent increment structure, which may not be satisfied in practice. The marginal rates model allows us to make a similar inference without assuming the Poisson assumption and the independent increment structure by modeling the mean rate function of recurrent events~\citep{10.1007/978-1-4684-6316-3_4, Lawless1995, Pepe01091993}. It is defined by
\begin{equation}
\label{marginalrates}
    E\{dN^*(t)\mid Z \}=d\mu(t\mid Z)=\exp(\beta_0Z)d\mu_0(t),
\end{equation} 
where $\mu(t|Z)$ is the mean rate function and  $\mu_0(\cdot)$ is an unknown continuous baseline mean rate function for recurrent events. The treatment effect is measured by $\beta_0$, the log-rate ratio between two treatment groups.\\

Let the solution to $U(\beta)=0$  with~\eqref{pscore} still be denoted by $\hat{\beta}$ for the marginal rates model. Lin et al. (2000) showed that it remains consistent even for the mean rates model.  On the other hand, the asymptotic variance of the estimator should be modified. To account for the with-subject dependence, a sandwich-type robust variance should be used for $Var(\hat{\beta})$. ~\citet{Lin2000} further justified that $n^{1/2}(\hat{\beta}-\beta_0)$ is asymptotically zero mean normal with covariance $A^{-1}\Sigma A^{-1}$, where
\begin{equation}
\label{fisherinf}
    A \equiv E\Bigl[\int_0^\tau (Z-\bar z(\beta_0,u))^2Y(u)exp(\beta_0Z)d\mu_0(t)\Bigl],
\end{equation}

\begin{equation}
\label{scorecov}
    \Sigma = E\Bigl[\int_0^\tau (Z_1-\bar{z}(\beta_0,u))dM_1(u)\int_0^\tau (Z_1-\bar{z}(\beta_0,v))dM_1(v)\Bigl],
\end{equation}
and $M_i(t)=N_i(t)-\int_0^tY_i(u)exp(\beta_0Z_i)d\mu_0(u)$. Note that for the standard AG model~\eqref{cox0} with the Poisson assumption, $\mu_0(t)=\Lambda_0(t)$ and $A$ in~\eqref{fisherinf0} agree with that in~\eqref{fisherinf}. Therefore, under the AG model, $A=\Sigma$ holds, and the covariance is reduced to $A^{-1}$ as argued. However, it does not hold in general for the mean rates model. The variance of $n^{1/2}(\hat{\beta}-\beta_0)$ is consistently estimated by  $\hat{A}^{-1}\hat{\Sigma} \hat{A}^{-1}$, where
\begin{equation*}
\label{fisherinf_emp}
    \hat{A} = \frac{1}{n}\sum_{i=1}^n \Bigl[\int_0^\tau \{Z_i-\bar Z(\hat{\beta},u)\}^2Y_i(u)exp(\hat{\beta}Z_i)d\hat{\mu}_0(t)\Bigl], 
\end{equation*}

\begin{equation*}
\label{scorecov_emp}
    \hat{\Sigma} = \frac{1}{n}\sum_{i=1}^n \Bigl[\int_0^\tau \{Z_i-\bar{Z}(\hat{\beta},u)\}d\hat{M}_i(u)
    \int_0^\tau \{Z_i-\bar{Z}(\hat{\beta},v)\}d\hat{M}_i(v)\Bigl],
\end{equation*}
where $\hat{M}_i(t)=N_i(t)-\int_0^tY_i(u)exp(\hat{\beta}Z_i)d\hat{\mu}_0(u)$ and 
\begin{eqnarray*}
\hat{\mu}_0(t)= \int_0^{\tau} \frac{d\sum_{i=1}^n N_i(u)}{nS^{(0)}(\hat{\beta},u)}.
\end{eqnarray*}
Denote $\hat{v}^2=n^{-1}\hat{A}^{-1}\hat{\Sigma} \hat{A}^{-1}$. Then, the robust Wald-type test can be made by the test statistic $\hat{\beta}/\hat{v}$ referring to the standard normal distribution under the null hypothesis $\beta_0=0$.  

\section{Event-driven type design for recurrent events}\label{sec3}

\subsection{Event-driven Design for the Standard AG Model}\label{sub31}

For the standard univariate survival analysis, in which only a single event is observed for each subject or the first event analysis in the context of the recurrent events, the event-driven design is widely used in practice. Instead of targeting the total number of subjects, the event-driven design focuses on the target number of events. Based on the Schoenfeld formula~\citep{Schoenfeld1981,Schoenfeld1983}, the target number of events in total for the two treatment groups is set, and the statistical analysis is conducted when the target number of events is observed.  In practice, the number of subjects should be specified. It is typically defined by specifying the accrual duration, follow-up duration, and survival function of the control group, in addition to the target hazard ratio. This procedure applies to the standard AG model for recurrent events with the Poisson assumption~\citep{PatriciaBernardo2001}.

Suppose the AG model~\eqref{cox0} holds. To obtain a simple power formula, an extra assumption, the censoring time $C$ is independent of $Z$ (see page 562 of Bernardo and Harrington 2001). Furthermore, Bernardo and Harrington (2001) supposed $\beta_0$ was small, which should be formally formulated by a local alternative $\beta_0=\delta/\sqrt{n}$, where $\delta$ is a constant.  Bernardo and Harrington (2001) showed that the asymptotic variance $A^{-1}$ of $\sqrt{n}(\hat{\beta}-\beta_0)$ is approximated by $\{\pi(1-\pi)L/n\}^{-1}$ or the variance of $\hat{\beta}$ is approximated by $\{\pi(1-\pi)L\}^{-1}$. Then, regarding $\beta_0$ is the target treatment effect to detect, one can set the target number of events so that the Wald-type test, or equivalently the logrank test, possesses the target power given the significance level. To be specific, suppose that a two-sided test will be performed with a significance level of $\alpha$ and a given statistical power $1-\gamma$. Let $z_{1-\alpha/2}$ and $z_{1-\gamma}$ be the $1-\alpha/2$ and $1-\gamma$ percentiles of the standard normal distribution, respectively. The required number of events in the two treatment groups in total is given by 
\begin{equation}
\label{schoenfeldformula}
    L=\frac{1}{\pi(1-\pi)}
   \left(\frac{z_{1 - \frac{\alpha}{2}} + z_{1-\gamma}}{\beta_0}\right)^{2}.
\end{equation}
For a balanced design, in which $\pi=1/2$, the required number of events is calculated as $4\cdot(z_{1-\alpha/2} + z_{1-\gamma})^2/\beta_0^2$.

This is an extension of the Schonfeld formula for the univariate survival analysis and enables us to conduct an event-driven study; likewise, the statistical analysis is conducted when the target number of events is observed. Note that the formula for the number of events does not rely on any quantities other than the treatment effect $\beta_0$, given the significance level and the target power. It implies that as long as the statistical analysis is conducted with the target number of events, the target power is ensured for the target treatment effect. \citet{PatriciaBernardo2001} proposed a method to set the number of subjects to enroll to obtain the target number of events within the planned duration of the study. Some strong assumptions on the probabilistic structure of the recurrent events are required for this calculation. However, it should be emphasized that misspecification of the probabilistic structure of the recurrent events may lead to prolongation of the study duration, but does not lead to underpowered studies. It is a very important feature since information on the probabilistic structure of the recurrent events at the design stage is very limited. 

\subsection{Existing Sample Size Planning Methods with the Robust Variance}\label{sub32}

On one hand, as argued in Section~\ref{sub31}, the Poisson-type assumption enables event-driven designs for recurrent events using the AG model. On the other hand, the Poisson-type assumption might be too strong because recurrent events are usually correlated within each subject, and the occurrence of a new event tends to be associated with previous events. In this subsection, we review methods for deriving the required number of events or subjects without the Poisson assumption, using robust variance to account for within-subject dependence in sample size planning.

\citet{Matsui2005} proposed the sample size calculation method for the logrank test for recurrent events. Although the test is nonparametric, determining the sample size requires specifying several parameters. \citet{Matsui2005} proposed specifying them based on the mixed Poisson process model. \citet{Song2008} proposed the sample size formula with the covariate-adjusted robust logrank test for the recurrent events. Their motivation was to provide a formula that did not rely on parametric models. Although their formula consists of only nonparametric quantities, they are hard to estimate at the design stage unless some individual participants' data from previously conducted studies under a similar design are available. Thus, the resulting sample size should be subject to uncertainty at the design stage. Following Song's derivation, ~\citet{Tang2019} also gave sample size formulas for the robust Wald test from the AG model for different study goals. To determine sample size, they provided the analytic expressions for the robust variance by assuming a Weibull event rate function or a piecewise constant rate function.   

All the methods introduced in the last paragraph proposed a formula to calculate the number of subjects. On the other hand, based on the robust score test, ~\citet{Ingel2014} proposed a formula to calculate the target number of events in the two treatment groups in total as
\begin{equation}
\label{schoenfeldformula2}
     L=  \frac{1}{\pi (1-\pi)}\left\{1 +Var(\lambda)\frac{\tau}{E(\lambda)}\right\}\cdot
         \left(\frac{z_{1 - \frac{\alpha}{2}} + z_{1-\gamma}}{\beta_0}\right)^{2},
\end{equation}
where $\lambda$ is an individual-specific random constant hazard over time, and the expectation and the variance are taken over the study population. The formula enables us to take an event-driven design for recurrent events. Comparing the formulas~\eqref{schoenfeldformula} and~\eqref{schoenfeldformula2},  one observes that the formula for the robust test~\eqref{schoenfeldformula2} is the formula under the Poisson assumption~\eqref{schoenfeldformula} multiplied by the inflation factor (IF), 
\begin{equation*}
\label{IF}
     IF= \left\{1 +Var(\lambda)\frac{\tau}{E(\lambda)}\right\}.
\end{equation*}
This inflation factor was derived based on the robust variance proposed ~\citet{Al-Khalidi2011}. Even assuming the constant hazard for each subject, it is impossible to specify $E(\lambda)$ and $Var(\lambda)$ without specifying some probabilistic structure of the recurrent events. Regarding the event process as a mixed Poisson process, given an unobservable subject-specific frailty $\eta_i$, the conditional intensity function of this subject may be modeled as 
\begin{equation}
\label{frailtymodel}
    \lambda_i\left( t \mid Z_i, \eta_i \right)=\eta_i\cdot \lambda_0(t)\cdot exp(\beta_0Z_i),
\end{equation}
where $\lambda_0(t)$ is a common baseline intensity function of the target population. The term $\eta_i$ is often called a frailty in survival analysis, indicating heterogeneity in the risk of event recurrence within the population. Typically, $\eta_i$ is assumed to follow a certain distribution with mean 1 and variance $\theta$~\citep{Balan2020}, with larger $\theta$ indicating higher heterogeneity. Under this frailty model, ~\citet{Ingel2014} proposed to use the following formula, generalizing (\ref{schoenfeldformula2}) to time-varying baseline mean rate functions, 
\begin{equation}
    \label{LL_ingel}
     L= \frac{1}{\pi (1-\pi)} \left\{1 + \theta\cdot\mu_0(\tau)\cdot
         \left(\frac{1 + \exp\bigl(2\beta_0\bigr)}{1 + \exp\bigl(\beta_0\bigr)}\right)\right\}\cdot
         \left(\frac{z_{1 - \frac{\alpha}{2}} + z_{1-\gamma}}{\beta_0}\right)^{2}.
\end{equation}

To determine the number of events $L$, one needs to specify the variance $\theta$ of the frailty term and the baseline mean rate function $\mu_0(t)$. For the baseline mean rates function, one may use the Weibull form function $\mu_0(t)=\lambda t^{\nu}$ for time-varying event rates. Misspecification of these quantities would lead to a poor estimate of the power. Thus, even if the statistical analysis is conducted when the target number of events calculated by~\eqref{LL_ingel} is attained, the power may be less than the target one. \citet{Ingel2014} demonstrated how to calculate the target number of subjects to observe the target number of events within the planned study duration, given the accrual and follow-up durations. Regarding accrual of subjects, a constant rate or uniform accrual pattern within an accrual period is typically assumed. For follow-up duration, several options are available: setting an equal follow-up duration for all subjects, considering only administrative censoring, or assuming an exponential distribution for the censoring time $C_i$. 

\subsection{The Proposed Procedure: Blinded Monitoring of $v^2$}\label{sub33}

Let us consider the power of the robust Wald test based on the mean rates model~\eqref{marginalrates}. 
Consider testing the null hypothesis $H_0:\beta_0=0$ against a local alternative $H_1:\beta_0=\tfrac{\delta}{\sqrt{n}}$, where $\delta$ is a constant. As argued in Section~\ref{sub22}, $\sqrt{n}(\hat{\beta}-\beta_0)$ approximately follows $N(0, A^{-1}\Sigma A^{-1})$ or the distribution of $\beta$ is approximated by $N(\beta_0, v^2)$, where $v^2=n^{-1} A^{-1} \Sigma A^{-1}$. Then,  for a given two-sided significance level $\alpha$, the power  is given by
\begin{equation}
\label{power}
    power=\Phi(-z_{1-\alpha/2}-\beta_0/v)+1-\Phi(z_{1-\alpha/2}-\beta_0/v),
\end{equation}
where $\Phi$ is the cumulative distribution function of the standard normal distribution. Thus, for the target treatment effect $\beta_0$, the robust Wald test can achieve the target power with an appropriately chosen $v^2$ or $n$. The resulting $v^2$ is denoted by $v_{\text{target}}^2$.

From Slutsky's theorem, the asymptotic distribution of the robust Wald test $\hat{\beta}/v$ does not change if $A$ and $\Sigma$ in $v^2$ are replaced with their consistent estimators. We consider consistent estimators under the local alternative. As argued in Section~\ref{sub22}, $\hat{A}$ has the same stochastic limit as $\pi(1-\pi)L/n$. In the Appendix, we show that the limit of $\hat{\Sigma}$ can also be written in a form free of treatment allocation $Z_i$ when $\pi=1/2$. 
Then, the variance $v^2$, under a 1:1 randomization ratio, can be estimated in a blinded manner by
\begin{equation}
\label{estv2}
    \hat{v}_{\text{blind}}^2=\left\{\frac{L}{4}\right\}^{-1}\cdot\left(\frac{1}{4}\sum_{i=1}^{n}\bigl[N_i(\tau)-\hat{\mu}_{0}(\tau \wedge C_i)\bigl]^2\right)\cdot \left\{\frac{L}{4}\right\}^{-1}.
\end{equation}

As the study proceeds, the maximum follow-up duration $\tau$ increases, as well as the number of events in the two treatment groups $L$. Thus, by continuing the follow-up, the target power can be obtained with a sample size chosen $n$. Thus, we propose to determine the timing of the statistical analysis by monitoring  $\hat{v}_{\text{blind}}^2$ by \eqref{estv2}. That is, the statistical analysis is conducted when $\hat{v}_{\text{blind}}^2$ attains the $v^2_{\text{target}}$ for the target power. To be more specific, the following procedure is proposed.

\begin{enumerate}[label=Step \arabic*]
    \item Set a two-sided significance level $\alpha$, target power, and the log-rate ratio $\beta_0$ as the minimum clinically relevant treatment effect.
    \item Make a (tentative) calculation of the number of subjects with some most likely assumptions of the probabilistic structure of the recurrent events. For example, one may set the target number of subjects $n$, assuming the mixed Poisson process as done by Ingel and Jahn-Eimermacher (2014), outlined in Section~\ref{sub32}.
    \item Calculate the target variance $v_{\text{target}}^2$ for the setting in Step 1 with the power formula \eqref{power}.
    \item Monitor $\hat{v}_{\text{blind}}^2$ in a blinded manner continuously as the study proceeds.
    \item Once $\hat{v}_{\text{blind}}^2\leq v_{\text{target}}^2$, conduct the final analysis using the marginal rates model. 
\end{enumerate}

The proposed procedure suggests choosing the timing of the final analysis adaptively by monitoring $\hat{v}_{\text{blind}}^2$ or, equivalently, the predicted power based on \eqref{power} with $v=\hat{v}_{\text{blind}}$ to attain the target power. In Step 2, we need to specify the number of subjects. One may use any existing method discussed in section~\ref{sub32}. However, these methods strongly rely on modeling of the recurrent events, and the true probabilistic structure is hardly known at the design stage. Our procedure protects underpowered studies with misspecification at the design stage. However, it may lead to substantial prolongation of the study if the supposed model at the design stage is substantially far from reality. Thus, one needs to specify the most likely models in Step 2.

To note, when the accumulated number of events is small or the follow-up duration $\tau$ is short, the resulting $\hat{v}^2_{\text{blind}}$ tends to be unstable. We consider initiating the monitoring process after either a preset time has elapsed or a predetermined threshold of events has been reached. While continuous monitoring of $\hat{v}_{\text{blind}}^2$ is ideal in Step 4, practical constraints, such as the impossibility of accumulating updated information daily due to limited human and financial resources, may exist. If this is the case, periodic monitoring on a weekly or monthly basis may be more practical. 

\section{Simulation studies}\label{sec4}

\subsection{Overview}\label{sub41}

The simulation studies consist of two settings. In the first setting, the sample sizes are calculated when the planning assumptions of the parameters are fulfilled. Then, we evaluate the proposed blind monitoring procedure in terms of the ability to achieve the target power and control the type 1 error probability. In the second setting, we set the sample sizes with some misspecified parameters at the design stage. The focus of the second setting is to test the ability of the proposed method to maintain statistical power when certain sample size planning assumptions are misspecified. We compare the performance with the event-driven design by Ingel and Jahn-Eimermacher (2014)~\cite{Ingel2014}. Although Ingel and Jahn-Eimermacher (2014) also proposed a blinded sample-size re-calculation approach to identify the quantities needed to calculate \eqref{LL_ingel}, we evaluate their method to determine the timing of the statistical analysis based on the specification at the design stage. Thus, we refer to the method as the fixed design. In our proposed monitoring procedure, we follow Step 2 to determine the sample size $n$. That is, the same $n$ was supposed as the fixed design. However, the timing of the statistical analysis was not determined by $L$, instead by monitoring $\hat{v}_{\text{blind}}^2$; the statistical analysis was conducted once $\hat{v}_{\text{blind}}^2$ agreed with $v_{\text{target}}^2$ of the target power. In Section~\ref{sub42}, we give details of the simulation settings and data generation of the simulation studies. In Section~\ref{sub43}, we summarize the results when the specification at the design stage is correct. In Section~\ref{sub44}, the results of misspecified cases are presented.

\subsection{Data Generation}\label{sub42}

We simulate a randomized controlled trial to compare the treatment effect between an experimental and a control group. All subjects are assigned randomly to either group by $Z_i \sim \text{Bin}(1, 0.5)$. We define a baseline mean rate function as a Weibull-type one, $\mu_0(t)=\lambda t^{\nu}$. The parameter $\nu$ describes the time trend of event rates: $\nu=1$ corresponds to a constant rate; $\nu<1$ implies that the event happens less frequently as time passes by, while $\nu>1$ means that the event rate increases with time. To account for the within-subject event times correlation, we applied the model \eqref{frailtymodel} with the frailty following a gamma distribution with both shape and rate parameters set to $\tfrac{1}{\theta}$ for the individual event rate. Event times are generated using an inverse transformation method and a recursive algorithm proposed by ~\citet{Jahn-Eimermacher2015}. Let all subjects enter the trial uniformly between 0 and $\tau_a$ years and are followed for at least $\tau_f$ years. That is, only administrative censoring is considered. We set $\beta_0=\log{(0.8)}$ when evaluating the power and $\beta_0=0$ when evaluating the type 1 error probability. We simulated 2000 datasets to calculate empirical power and type 1 error rate. To determine the target $L$ by the method of Ingel et al. and $v_{\text{target}}^2$ for the fixed and proposed methods, respectively, we set the target power as 0.8 and a two-sided type 1 error rate of 0.05.

In the first setting of the simulation study, when the sample size planning assumptions are correctly specified at the design stage, each parameter combination $\{\lambda, \nu, \theta, (\tau_a, \tau_f)\}$ is used to calculate the sample size $n$ for both designs and the target number of events $L$ for the fixed design. The same parameters will be used to generate datasets. Details of parameter settings are specified in Table~\ref{tab:parametercombi}.

In the second setting of the simulation study, we considered three scenarios of planning assumptions misspecifications, which might not fit the model to generate datasets, or be misspecified. To clarify, in each scenario, a true parameter is used for data generation while an assumed parameter is used to determine the sample size at the design stage. In scenario-1, we evaluated the impacts of ignoring frailty variance when planning sample sizes. we generate the datasets with the frailty variance $\theta_{\text{true}}=0, 0.1, 0.2, 0.3$, but $(n, L)$ is calculated using $\theta_{\text{assumed}}=0$ at the design stage. In scenario-2, similar to the first case, the research team may have considered estimating heterogeneity but ended up misspecifying it. The datasets are generated with $\theta_{\text{true}}=0.5$ while $(n, L)$ is determined using $\theta_{\text{assumed}}=0.3, 0.4, 0.5, 0.6, 0.7$. In scenario-3, the impacts of a misspecified baseline mean rate function were investigated. To this end, the data were generated from the mixed Poisson process with the Weibull-type baseline mean rate function of $\nu_{true}=0.5, 1, 2$. At the design stage, we determined $(n, L)$ with $\nu_{\text{assumed}} = 1$, which corresponds to a constant rate function $\lambda_0 = 1$.

\subsection{Validity of the proposed monitoring procedure}\label{sub43}

Table~\ref{tab:empirical} summarizes the empirical power, type 1 error rate, as well as the distribution of the analysis time. For the fixed design, the analysis time is measured by the calendar years when $L$ events are observed. Here, the calendar years imply years from the enrollment of the first subject into the study. For the proposed monitoring procedure, aside from the calendar years when $v_{\text{target}}^2$ is attained, we also reported the corresponding number of events at the final analysis. In general, when the sample size planning assumptions were met, the empirical power of the proposed method was close to 0.8 and comparable to that of the fixed design. Likewise, the type 1 error probability using the proposed monitoring procedure remained close to the nominal level of 0.05. In most cases, the proposed method yielded a similar analysis time to the fixed design, around 3 years. Exceptions happened when the baseline mean rate function increased with time, and the heterogeneity became larger. The statistical power of the fixed design dropped substantially under $\nu=2$ and $\theta>0$, remaining roughly 2–10 points under the target power level. The results might suggest potential violations of the underlying assumptions to derive the inflation factor in \eqref{LL_ingel}, such as unequal follow-up durations and significant variability in individual event rates~\citep{Al-Khalidi2011}. As a result, the sample size might be underestimated. In contrast, the proposed method maintained the power close to 0.8. However, it prolonged the analysis time compared to the fixed design: the median analysis time extended up to 3.96 years (Q1, Q3: 3.61, 4.51), with a corresponding median event count twice that of the $L$ derived at the design stage (proposed: 6541 vs. fixed: 3123). To note, in the last three scenarios, in which the mean rate function increases with time, and the frailty variance is large, some simulated datasets did not reach the $v_{\text{target}}^2$ of the proposed method even after 15 years from the study initiation. We evaluated empirical type 1 error rates and powers by removing such datasets. The maximum rate of removal was 2.95\% with scenario $\{\nu=2, \theta=1, (\tau_a, \tau_f)=(1,2)\}$ when calculating the empirical power.  

To conclude, the results from different event-generating processes and heterogeneity justified the use of $\hat{v}_{\text{blind}}^2$ in the form of~\eqref{estv2} to monitor the predicted power of clinical trials with recurrent events. With sample sizes derived from correctly specified parameters at the design stage, the proposed method achieved the desirable level of power, albeit with a slightly longer trial duration, compared to the fixed design. Moreover, the proposed method also showed robustness to large variability in event rates, follow-up durations, and susceptibility to the events of interest.

\subsection{Performance of the proposed monitoring procedure under parameter misspecification at the design stage}\label{sub44}

Table~\ref{tab:misspecified} shows the power and analysis time of the fixed and our proposed method under the three scenarios of misspecification we introduced in Section~\ref{sub42}. In scenario-1, under the fixed design, as $\theta_{true}$ increased, we observed a decrease in power from 0.801 to 0.562. In contrast, by applying the proposed blind monitoring of the robust variance, the empirical power was maintained near 0.8 at the cost of prolonging the total follow-up period, with the longest median analysis time being 8.61 years (Q1, Q3: 7.15, 10.65). In scenario-2, with $\theta_{\text{assumed}} = 0.3$ or $0.4$ (underestimated case), the power under the fixed design decreased by up to 10\% while the proposed method recovered the power around 0.8. With $\theta_{\text{assumed}} = 0.6$ or $0.7$ (overestimated case), the target $L$ was still achieved at 3 years, but resulted in increased power under the fixed design; notably, the proposed blind monitoring procedure mitigated the overpowering with the median analysis time about 2.58 years or less, along with a smaller number of events when conducting the final analysis. The results of Scenario-3 showed that the misspecification of the baseline mean rate function would affect the trial durations under both the fixed design and the proposed method. Nevertheless, by monitoring the robust variance and thus the predicted power, we controlled the power closer to 0.8 compared with its counterpart under the fixed design.

In summary, the proposed method can protect the statistical power of clinical trials with recurrent events from not only underestimation but also overestimation of sample sizes. Based on the target value of the robust variance derived at the design stage, the proposed monitoring procedure minimizes the impacts of misspecified planning assumptions by adjusting the follow-up durations.

\section{Example}\label{sec5}

In this section, we illustrate the proposed blind monitoring procedure to determine the timing of final analysis using chronic granulomatous disease (CGD) data~\citep{Fleming2005}, which is available in the \textit{R} package \textit{SURVIVAL}. The International Chronic Granulomatous Disease Cooperative Study Group designed a multicenter, randomized, double-blind trial to evaluate the efficacy of interferon gamma in preventing severe infections in CGD patients. In the original study design, subjects were randomized to either the placebo or the interferon gamma group to receive a 12-month treatment. The primary endpoint was the time to the first severe infection, and multiple times to serious infection were also recorded.~\citep{doi:10.1056/NEJM199102213240801}. The trial was terminated early based on the results of an interim analysis in July 1989~\citep{Fleming2005}. A total of 128 eligible patients were enrolled in the trial, with 65 assigned to the placebo group and 63 to the interferon gamma group~\citep{Therneau2024}. 

We use multiple times to serious infections as the recurrent events of interest for illustration. We explain the procedure following the steps in section~\ref{sub33}. In Step 1 (see Section~\ref{sub33}), we set the two-sided significance level $\alpha=0.05$ and the target power $1-\gamma=0.8$. used the same target power and significance level as in our simulation studies. The target treatment effect was set as $\beta_0$ to be $\log(0.3)$. Following Ingel and Jahn-Elimermacher (2012), we use the mixed Poisson model to determine the sample size in Step 2. In reality, we face difficulty in setting relevant parameters in the mixed Poisson model at the design stage, and may use some existing datasets to this end. Here, for illustrative purposes, we set values for parameters in the model properly fitting the placebo group of the CGD data. As described above, the planned follow-up duration $\tau$ was 1 year. Assuming a constant baseline intensity, the estimated $\hat{\lambda}_0$ was 1.10 events/(person $\cdot$ year). Applying an EM algorithm~\citep{Balan2022} to the dataset of the placebo group, we estimated the frailty variance $\hat{\theta}$ to be 0.825. With a target power of 0.8 and a two-sided significance level of 0.05, we derived the target $L=39$ for the fixed design. In this illustration, we set $n = 128$. That is, we would conduct the final analysis if we collected 39 events from the 128 patients in the fixed design.

In Step 2, we do not follow the procedure in Section~\ref{sub33}. Instead, we also set $n = 128$ for our proposed procedure. In Step 3, we obtained the target variance $v_\text{target}^2= 0.185$ for the blind monitoring procedure. The dataset contains the date of randomization of each subject. In this illustration, the first date of randomization was regarded as the date of initiation of the study. The trial lasted from August 28, 1988, to the final follow-up date of January 17, 1990, covering 507 days. Monitoring began one month after the trial was initiated, and we evaluated three intervals to monitor the target variance $\hat{v}_{\text{target}}^2$ or the corresponding predicted power: daily (continuous), weekly, and monthly in Step 4. At the first point when the predicted power exceeded 0.8 (as in Step 5), we fitted the marginal rates model to estimate $\beta_0$ along with its robust standard error and corresponding \textit{p}-value from the robust Wald test.

An aspect of applying our proposed method is that the quantity $\hat{v}_\text{blind}^2$ or the resulting predicted power is an estimate and then subject to uncertainty in the estimation. To facilitate the decision to final analysis accounting for such uncertainty, we applied a bootstrap approach~\citep{Gonzalez2010} to compute confidence intervals for $\hat{v}^2$ during monitoring. For each resampled dataset, compute $\hat{v}^2$ at each monitoring time point $\tau, \tau \in \{\tau \in \mathbb{Z} \mid 30 \leq t \leq 507\}$. With 2,000 bootstrap samples, we obtained the distribution of $\hat{v}^2$ at each $\tau$, and formed the 95\% confidence intervals using the percentile approach.

In Table~\ref{tab:application}, we present the results of the statistical analysis based on the fixed design and the monitoring procedure. In the fixed design, we conducted the final analysis with the first $L=39$ events observed. It was on day 309 from the start of the study. The rate-ratio was estimated as 0.272 (95$\%$ CI: $0.114-0.645$) with $P=0.005$. The proposed continuous (daily) monitoring procedure suggests conducting the final analysis a little earlier on day 281 with the predicted power of 0.802. The rate-ratio was estimated as 0.294 (95$\%$ CI: $0.124-0.699$) with $P=0.0006$. If we follow the weekly or the monthly monitoring procedure, the final analyses were on day 287 and day 307, respectively. 

\begin{figure}[htbp]
  \centering
  \includegraphics[width=1\textwidth]{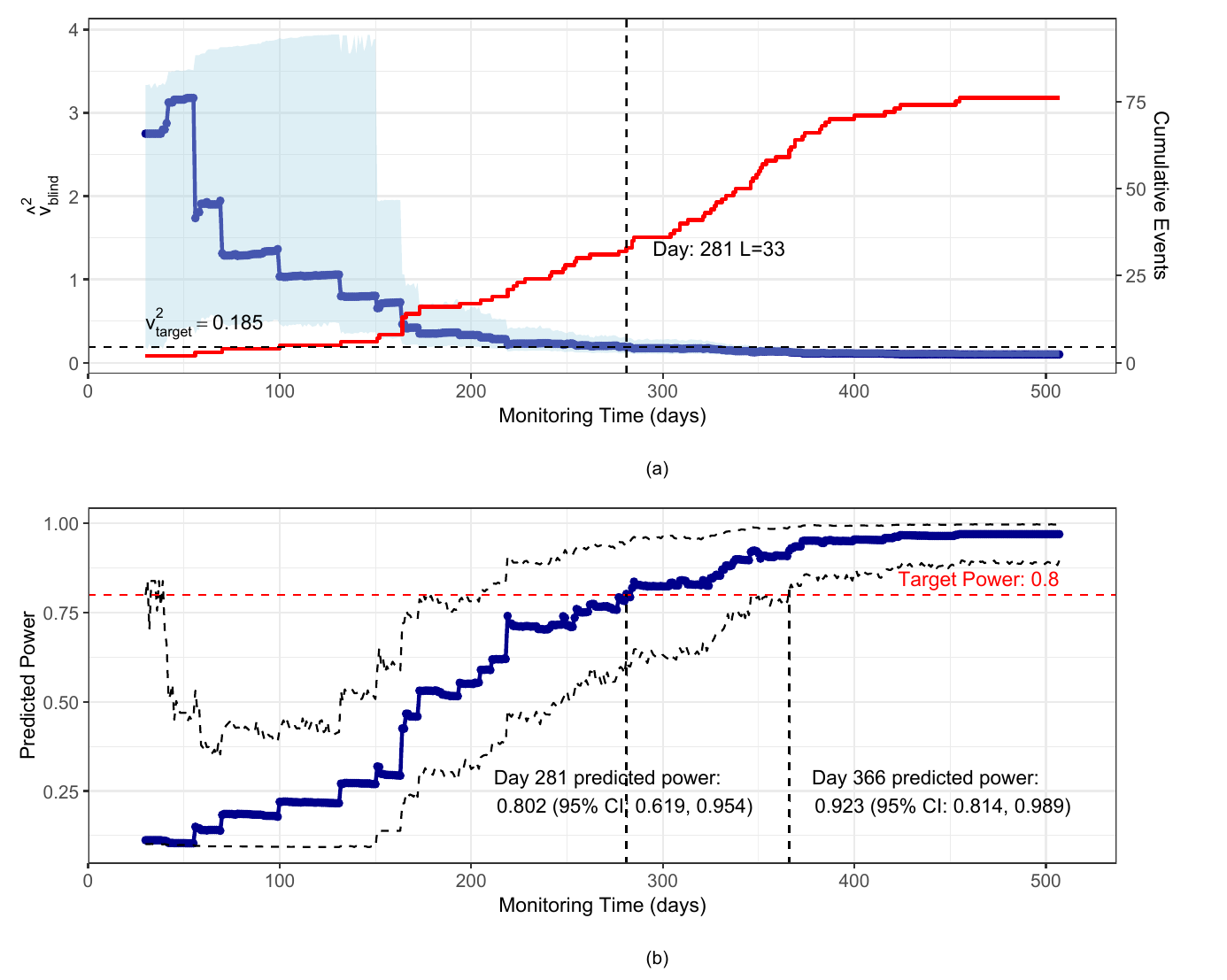}
  \caption{(a) shows the changes in $\hat{v}^2$ (dark blue curve) over monitoring time, with the light blue shaded region representing the bootstrapped confidence interval for $\hat{v}^2$. The red curve corresponds to the changes in event counts over time. (b) shows the transformed predicted power and corresponding bootstrapped upper and lower bounds using equation \eqref{power} with $\beta_0=\log(0.3)$ and two-sided significance level $\alpha=0.05$ over monitoring time.}
  \label{fig:overall}
\end{figure}

In Figure~\ref{fig:overall} (a), we present the cumulative number of events $L$ in both groups over the calendar time (the curve in red) and the estimated target variance $\hat{v}_{\text{target}}^2$ (the dark blue curve) with pointwise bootstrap 95$\%$ confidence intervals (the light-blue shaded areas). Figure~\ref{fig:overall} (b) shows the corresponding predicted power over the calendar time. Bootstrap confidence intervals for $\hat{v}_{\text{target}}^2$ or the predicted power would be useful for deciding the timing of the statistical analysis. For example, narrow CIs indicate that the estimate is less sensitive to fluctuations in the data. In Figure~\ref{fig:overall} (a), the black dashed line shows that the predicted power attained the target power of 0.8 at day 281 with continuous monitoring with 33 events. Alternatively, we converted $\hat{v}_\text{blind}^2$ to predicted power in Figure~\ref{fig:overall} (b). Specifically, we estimated the power 0.802 (95\% CI: $0.619-0.954$) on day 281 for continuous monitoring. One may be concerned with the rather lower predicted power of 0.619 at the lower bound of its 95\% confidence intervals, and then may hope to take a safer approach to attain enough statistical power. We could extend the follow-up until day 366. At this point, with the occurrence of 67 events, the lower bound of the predicted power exceeded 0.8, and the point estimate reached 0.923 (95\% CI: $0.814-0.989$). 

These results demonstrate that the monitoring procedure allows for an earlier evaluation, potentially reducing trial duration without compromising the accuracy of the treatment effect estimate. Thus, compared to the conventional fixed design, the monitoring procedure offers higher operational efficiency, greater flexibility, and timely decision-making.

\section{Discussion}\label{sec6}

In this paper, we proposed an event-driven type design for clinical trials with recurrent events. Instead of taking a purely event-driven approach, we monitored the sandwich variance under blinded treatment allocation and determined the timing of the statistical analysis when the monitored variance reached the target variance. This procedure ensures the target power regardless of any nuisance quantities other than the target treatment effect. To the best of our knowledge, this is the first realization of an event-driven type design for recurrent events that explicitly accounts for within-subject dependence through blinded monitoring of the sandwich variance.

A monitoring strategy based on the monitored robust variance does not remove the requirement to choose a sample size at the design stage. If the planned sample size is too small, the time required for the monitored variance to reach the target can become substantially longer than expected. Indeed, as observed in our simulation study with substantial misspecification of certain nuisance quantities, such as the baseline rate and frailty variance at the design stage, it might lead to noticeable prolongation of the trials. If this is the case, the trial may be infeasible to complete. Even if feasible, it may cause unacceptable delays in making scientific findings shared in public. Thus, it is essential to set an appropriate initial sample size that well reflects the true probabilistic structure of the recurrent events. Many authors developed methods to set sample sizes at the design stage, modeling the within-subject dependence, such as the mixed Poisson model~\citep{Cook1995,Tang2019,Ingel2014}. Any method can be applied in Step 2 of our proposed method. If the specification of the model at the design stage is suggested to be inappropriate during the study, one may hope to extend the sample size to avoid unacceptable prolongation of the study. It would be warranted to investigate how to identify the inappropriateness of the supposed model and expand the sample size under blind monitoring of data.

\bibliographystyle{unsrtnat}
\bibliography{references}  %%% Uncomment this line and comment out the ``thebibliography'' section below to use the external .bib file (using bibtex) .





% \end{thebibliography}

\clearpage

\appendix
\section*{Derivation of $\hat{v}^2_{\text{blind}}$}

In this appendix, we present how we approximated $\hat{A}$ and $\hat{\Sigma}$, respectively. Note that when $\beta_0 \to 0$, $\bar{Z}(\beta_0,t) = S^{(1)}(\beta_0,t)/S^{(0)}(\beta_0,t)$ approaches $\pi$~\citep{Schoenfeld1983}, and $(Z-\pi)=\pi(1-\pi)$. %may need to be verified, check conditions.
We assume independent censoring (condition 1) and $C \perp Z$ (condition 2). The score function is given by

\begin{equation}
    U(\beta_0) = \sum_{i=1}^n\int_{0}^\tau\{Z_i - \bar{Z}(\beta_0,u)\}(dN_i(t) - Y_i(u)\exp(\beta_0Z_i)d\mu_{0}(u)).
\end{equation}

\citet{Lin2000} proved that $n^{-1/2}U(\beta_0)$ converges in distribution to a normal distribution with mean 0 and variance $\Sigma$ as defined in~\eqref{scorecov}. Under the local alternative $H_1: \beta_0 = \frac{\delta}{\sqrt{n}}$, $\bar{z}(\beta_0,u)=\pi$ and
\begin{align}
    \Sigma &= E\bigl[(\int_0^{\tau}\{Z - \pi\}\{dN(u)-Y(u)d\mu_0(u)\})^2 \bigr]\\
    &= E\bigl[\{Z- \pi\}^2\{N(\tau)-\mu_0(\tau \wedge C)\}^2\bigr].
\end{align}
When $\pi=1/2$, it is represented as
\begin{align}
    \frac{1}{4}E\bigl[\{N(\tau)-\mu_0(\tau \wedge C)\}^2\bigr],
\end{align}
By replacing the expectation with the sample average and the theoretical quantities with their empirical counterparts, coupled with the fact that $A = \lim_{n \to \infty}\pi(1-\pi)\frac{L}{n}$, ~\eqref{estv2} is obtained.

\clearpage

\begin{table}[htbp]
  \centering
  \caption{Parameter settings of the simulation study to evaluate the operating characteristics of the proposed monitoring procedure under different event-generating processes}
    \begin{tabular}{lll}
    \hline
    Scale of the Weibull-form rate & $\lambda$ & 1 \\
    Shape of the Weibull-form rate & $\nu$    & 0.5, 1, 2 \\
    Frailty variance & $\theta$ & 0, 0.5, 1 \\
    Accrual and follow-up duration\textsuperscript{(a)} & $(\tau_{a}, \tau_{f})$ & (1, 2), (1.5, 1.5), (1, 2) \\
    \hline
    \end{tabular}%

    \vspace{2pt} % thin white-space to separate table and note
  \footnotesize
  \textsuperscript{(a)} Subjects are accrued uniformly in $[0,\tau_{a} ]$ years and followed at least $\tau_f$ years.
  \label{tab:parametercombi}%
\end{table}%

\begin{table}[htbp]
  \centering
  \caption{The operating characteristics of the proposed monitoring procedure compared with a fixed event-driven design when sample size planning assumptions are correctly specified at the design stage}
  \setlength{\tabcolsep}{4pt}
    \begin{adjustbox}{max width=\textwidth, max totalheight=\textheight-5cm, keepaspectratio}
    \begin{tabular}{llllllllrllll}
    \hline
          &       &       &       &       & \multicolumn{3}{c}{\textbf{Fixed}} &       & \multicolumn{4}{c}{\textbf{Proposed}} \\
\cline{6-8}\cline{10-13}          &       &       &       &       &       &       & \multicolumn{1}{c}{\textbf{Analysis time: Median (Q1, Q3)}} &       &       &       & \multicolumn{2}{c}{\textbf{Analysis time: Median (Q1, Q3)}} \\
\cline{8-8}\cline{12-13}    \textbf{$\nu$} & \textbf{$\theta$} & \textbf{$(\tau_{a},\tau_{f})$} & \textbf{n}\textsuperscript{(a)}  & \textbf{L}\textsuperscript{(a)}  & \textbf{Power} & \textbf{Type I error} & \textbf{Calendar time (years)} &       & \textbf{Power} & \textbf{Type I error} & \textbf{Calendar time (years)} & \textbf{\# of events} \\
    \hline
    0.5   & 0     & (1, 2) & 445   & 631   & 0.796  & 0.047  & 2.98 (2.85, 3.12) &       & 0.795  & 0.050  & 3.04 (2.76, 3.33) & 637 (609, 667) \\
          &       & (1.5, 1.5) & 470   & 631   & 0.795  & 0.049  & 3.00 (2.88, 3.12) &       & 0.799  & 0.049  & 3.03 (2.81, 3.29) & 636 (608, 666) \\
          &       & (2, 1) & 502   & 631   & 0.792  & 0.053  & 3.00 (2.89, 3.10) &       & 0.791  & 0.048  & 3.02 (2.84, 3.23) & 634 (610, 663) \\
          & 0.5   & (1, 2) & 764   & 1085  & 0.818  & 0.048  & 3.01 (2.87, 3.14) &       & 0.811  & 0.052  & 3.02 (2.66, 3.46) & 1088 (1009, 1173) \\
          &       & (1.5, 1.5) & 790   & 1061  & 0.797  & 0.056  & 3.00 (2.89, 3.12) &       & 0.802  & 0.056  & 3.05 (2.74, 3.43) & 1074 (1002, 1156) \\
          &       & (2, 1) & 822   & 1034  & 0.791  & 0.048  & 3.00 (2.90, 3.10) &       & 0.791  & 0.048  & 3.05 (2.81, 3.34) & 1046 (984, 1117) \\
          & 1     & (1, 2) & 1084  & 1539  & 0.801  & 0.043  & 3.00 (2.87, 3.14) &       & 0.802  & 0.051  & 2.97 (2.52, 3.62) & 1530 (1388, 1712) \\
          &       & (1.5, 1.5) & 1109  & 1490  & 0.803  & 0.060  & 3.00 (2.88, 3.12) &       & 0.799  & 0.058  & 3.05 (2.68, 3.55) & 1510 (1380, 1662) \\
          &       & (2, 1) & 1141  & 1436  & 0.808  & 0.062  & 3.00 (2.90, 3.11) &       & 0.805  & 0.060  & 3.10 (2.79, 3.51) & 1476 (1354, 1606) \\
    1     & 0     & (1, 2) & 281   & 631   & 0.801  & 0.042  & 3.00 (2.93, 3.06) &       & 0.802  & 0.044  & 3.03 (2.87, 3.20) & 640 (604, 679) \\
          &       & (1.5, 1.5) & 312   & 631   & 0.796  & 0.052  & 2.99 (2.93, 3.06) &       & 0.797  & 0.052  & 3.02 (2.89, 3.17) & 639 (604, 676) \\
          &       & (2, 1) & 351   & 631   & 0.782  & 0.048  & 2.99 (2.94, 3.05) &       & 0.793  & 0.048  & 3.02 (2.90, 3.14) & 638 (605, 674) \\
          & 0.5   & (1, 2) & 600   & 1350  & 0.808  & 0.054  & 3.00 (2.93, 3.07) &       & 0.814  & 0.055  & 3.05 (2.81, 3.35) & 1374 (1247, 1538) \\
          &       & (1.5, 1.5) & 632   & 1278  & 0.788  & 0.048  & 3.00 (2.94, 3.06) &       & 0.795  & 0.051  & 3.11 (2.87, 3.34) & 1336 (1205, 1466) \\
          &       & (2, 1) & 670   & 1206  & 0.788  & 0.048  & 3.00 (2.94, 3.05) &       & 0.799  & 0.052  & 3.15 (2.96, 3.35) & 1296 (1184, 1419) \\
          & 1     & (1, 2) & 920   & 2069  & 0.797  & 0.052  & 3.00 (2.93, 3.06) &       & 0.806  & 0.052  & 3.09 (2.73, 3.50) & 2130 (1855, 2488) \\
          &       & (1.5, 1.5) & 951   & 1925  & 0.796  & 0.049  & 3.00 (2.94, 3.07) &       & 0.810  & 0.051  & 3.14 (2.86, 3.50) & 2042 (1805, 2357) \\
          &       & (2, 1) & 990   & 1781  & 0.783  & 0.044  & 2.99 (2.94, 3.05) &       & 0.792  & 0.048  & 3.24 (3.00, 3.56) & 1991 (1786, 2292) \\
    2     & 0     & (1, 2) & 111   & 631   & 0.780  & 0.056  & 3.00 (2.96, 3.04) &       & 0.782  & 0.058  & 3.05 (2.91, 3.18) & 656 (593, 726) \\
          &       & (1.5, 1.5) & 134   & 631   & 0.781  & 0.056  & 3.00 (2.95, 3.04) &       & 0.792  & 0.060  & 3.04 (2.92, 3.16) & 652 (594, 714) \\
          &       & (2, 1) & 162   & 631   & 0.788  & 0.050  & 3.00 (2.96, 3.04) &       & 0.791  & 0.052  & 3.04 (2.93, 3.15) & 651 (592, 713) \\
          & 0.5   & (1, 2) & 431   & 2452  & 0.781  & 0.050  & 3.00 (2.96, 3.03) &       & 0.795  & 0.051  & 3.17 (2.90, 3.55) & 2786 (2250, 3626) \\
          &       & (1.5, 1.5) & 454   & 2141  & 0.758  & 0.058  & 3.00 (2.96, 3.03) &       & 0.793  & 0.059  & 3.32 (3.08, 3.65) & 2771 (2286, 3474) \\
          &       & (2, 1) & 482   & 1877  & 0.749  & 0.048  & 3.00 (2.96, 3.04) &       & 0.812  & 0.049  & 3.53 (3.30, 3.87) & 2922 (2419, 3689) \\
          & 1     & (1, 2) & 750   & 4273  & 0.799  & 0.052  & 3.00 (2.97, 3.04) &       & 0.786* & 0.052  & 3.29 (2.90, 3.88) & 5329 (3912, 7814) \\
          &       & (1.5, 1.5) & 773   & 3650  & 0.757  & 0.050  & 3.00 (2.96, 3.04) &       & 0.788* & 0.048  & 3.65 (3.26, 4.25) & 6000 (4500, 8585) \\
          &       & (2, 1) & 801   & 3123  & 0.711  & 0.046  & 3.00 (2.97, 3.04) &       & 0.797* & 0.048  & 3.96 (3.61, 4.51) & 6541 (5097, 9098) \\
    \hline
    \end{tabular}
     \end{adjustbox}

         \vspace{2pt} 
  \raggedright\footnotesize
  \textit{Note}: Power is evaluated using 2000 simulated trials. Type 1 error rate is evaluated using another 2000 simulated trials under $\beta_0 = 0$. \\
  \textsuperscript{(a)} The sample size \textit{n} and the number of events \textit{L} under the fixed design are calculated with parameters $\{\nu, \theta, (\tau_{a}, \tau_{f})\}$ to achieve target power 0.8 under a treatment effect  $\beta_0 = 0.8$ and a two-sided significance level 0.05.\\
  * indicates potential underestimated empirical power due to simulated datasets that did not fulfill the stopping criteria in step 5.
  \label{tab:empirical}%
\end{table}%

\begin{table}[htbp]
  \centering
  \caption{The empirical power and analysis timing of the proposed monitoring procedure compared with a fixed event-driven design when sample size planning assumptions are misspecified at the design stage}
  \setlength{\tabcolsep}{4pt}
    \begin{adjustbox}{max width=\textwidth, max totalheight=\textheight-5cm, keepaspectratio}
    
    \begin{tabular}{llllllllll}
    \hline
    \rowcolor[rgb]{ .816,  .816,  .816} Scenario-1 &       &       &       &       &       &       &       &       &  \\
    \hline
          &       &       &       & \multicolumn{2}{c}{\textbf{Fixed}} &       & \multicolumn{3}{c}{\textbf{Proposed}} \\
\cline{5-6}\cline{8-10}          &       &       &       &       & \multicolumn{1}{c}{\textbf{Analysis time: Median (Q1, Q3)}} &       &       & \multicolumn{2}{c}{\textbf{Analysis time: Median (Q1, Q3)}} \\
\cline{6-6}\cline{9-10}    \textbf{$\theta_{\text{true}}$} & \textbf{$\theta_{\text{assumed}}$} & \textbf{n}\textsuperscript{(a)}  & \textbf{L}\textsuperscript{(a)}  & \textbf{Power} & \textbf{Calendar time (years)} &       & \textbf{Power} & \textbf{Calendar time (years)} & \textbf{\# of events} \\
    \hline
    0\textsuperscript{(b)}  & \multirow{4}[2]{*}{0} & \multirow{4}[2]{*}{281} & \multirow{4}[2]{*}{631} & 0.801  & 3.00 (2.93, 3.06) &       & 0.802 & 3.03 (2.87, 3.20) & 641 (604, 678) \\
    0.1   &       &       &       & 0.721  & 2.99 (2.93, 3.07) &       & 0.799 & 3.79 (3.53, 4.08) & 829 (768, 903) \\
    0.2   &       &       &       & 0.641  & 2.99 (2.91, 3.08) &       & 0.814  & 5.18 (4.66, 5.79) & 1181 (1062, 1339) \\
    0.3   &       &       &       & 0.562  & 3.00 (2.91, 3.09) &       & 0.786 & 8.61 (7.15, 10.65) & 2033 (1680, 2542) \\
    \hline
    \rowcolor[rgb]{ .816,  .816,  .816} Scenario-2 &       &       &       &       &       &       &       &       &  \\
    \hline
          &       &       &       & \multicolumn{2}{c}{\textbf{Fixed}} &       & \multicolumn{3}{c}{\textbf{Proposed}} \\
\cline{5-6}\cline{8-10}          &       &       &       &       & \multicolumn{1}{c}{\textbf{Analysis time: Median (Q1, Q3)}} &       &       & \multicolumn{2}{c}{\textbf{Analysis time: Median (Q1, Q3)}} \\
\cline{6-6}\cline{9-10}    \textbf{$\theta_{\text{true}}$} & \textbf{$\theta_{\text{assumed}}$} & \textbf{n}\textsuperscript{(a)}  & \textbf{L}\textsuperscript{(a)}  & \textbf{Power} & \textbf{Calendar time (years)} &       & \textbf{Power} & \textbf{Calendar time (years)} & \textbf{\# of events} \\
    \hline
    \multirow{5}[2]{*}{0.5} & 0.3   & 473   & 1063  & 0.711  & 3.00 (2.93, 3.08) &       & 0.807 & 5.16 (4.48, 6.04) & 1975 (1687, 2354) \\
          & 0.4   & 536   & 1206  & 0.753  & 3.00 (2.93, 3.07) &       & 0.795 & 3.83 (3.43, 4.29) & 1603 (1420, 1826) \\
          & 0.5\textsuperscript{(b)} & 600   & 1350  & 0.808  & 3.00 (2.93, 3.07) &       & 0.814 & 3.05 (2.81, 3.35) & 1374 (1247, 1538) \\
          & 0.6   & 664   & 1494  & 0.840  & 3.00 (2.94, 3.06) &       & 0.799 & 2.58 (2.40, 2.79) & 1243 (1134, 1368) \\
          & 0.7   & 728   & 1638  & 0.865  & 3.00 (2.94, 3.07) &       & 0.794 & 2.25 (2.12, 2.41) & 1145 (1067, 1243) \\
    \hline
    \rowcolor[rgb]{ .816,  .816,  .816} Scenario-3 &       &       &       &       &       &       &       &       &  \\
    \hline
          &       &       &       & \multicolumn{2}{c}{\textbf{Fixed}} &       & \multicolumn{3}{c}{\textbf{Proposed}} \\
\cline{5-6}\cline{8-10}          &       &       &       &       & \multicolumn{1}{c}{\textbf{Analysis time: Median (Q1, Q3)}} &       &       & \multicolumn{2}{c}{\textbf{Analysis time: Median (Q1, Q3)}} \\
\cline{6-6}\cline{9-10}    \textbf{$\nu_{\text{true}}$} & \textbf{$\nu_{\text{assumed}}$} & \textbf{n}\textsuperscript{(a)}  & \textbf{L}\textsuperscript{(a)}  & \textbf{Power} & \textbf{Calendar time (years)} &       & \textbf{Power} & \textbf{Calendar time (years)} & \textbf{\# of events} \\
    \hline
    0.5   & \multirow{3}[2]{*}{1} & \multirow{3}[2]{*}{600} & \multirow{3}[2]{*}{1350} & 0.791  & 6.76 (6.43, 7.09) &       & 0.793  & 6.77 (5.63, 8.36) & 1354 (1225, 1511) \\
    1\textsuperscript{(b)}  &       &       &       & 0.808  & 3.00 (2.93, 3.07) &       & 0.814 & 3.05 (2.81, 3.35) & 1374 (1247, 1538) \\
    2     &       &       &       & 0.772  & 2.05 (2.03, 2.08) &       & 0.797  & 2.18 (2.08, 2.28) & 1560 (1391, 1767) \\
    \hline
    \end{tabular}
    \end{adjustbox}

     \vspace{2pt} 
  \raggedright\footnotesize
  \textit{Note}:  power is evaluated using 2000 simulated trials under $\beta_0=log(0.8)$.\\
\textsuperscript{(a)} The sample size n and the number of events for the fixed design are calculated using $\{\nu = 1, \theta_{\text{assumed}}, (\tau_{a}, \tau_{f}) = (1, 2) \}$  under scenario-1 and scenario-2 or $\{\nu_{\text{assumed}}, \theta = 0.5, (\tau_{a}, \tau_{f}) = (1, 2)\}$  under scenario-3 with $\beta_0=log(0.8)$, and allocation ratio 1:1 for a target power 0.8 and a two-sided significance level 0.05.\\
\textsuperscript{(b)} The situation in which all sample size planning assumptions are correctly specified.\\

  \label{tab:misspecified}%
\end{table}%

\begin{table}[htbp]
  \centering
  \caption{Results of CGD data under $\beta_0=log(0.3)$ for a target power of 0.8 and two-sided significance level 0.05: fixed design vs. proposed monitoring procedure}
    \begin{tabular}{lllll}
    \hline
          & \multirow{2}[4]{*}{\textbf{Fixed}} & \multicolumn{3}{c}{\textbf{Proposed}} \\
\cmidrule{3-5}          &       & \textbf{Continuous} & \textbf{Weely} & \textbf{Monthly} \\
    \hline
    Calendar time (days) & 309   & 281   & 287   & 307 \\
    \# of events & 39    & 33    & 36    & 38 \\
    Predicted power\textsuperscript{(a)} & NA    & 0.802  & 0.827  & 0.826  \\
    $\hat{\beta}$ (SE) & -1.304 (0.441) & -1.224 (0.441) & -1.177 (0.448) & -1.221 (0.442) \\
    $\exp{(\hat{\beta})}$ (95\% CI) & 0.272 (0.114, 0.645) & 0.294  (0.124, 0.699) & 0.308 (0.128, 0.742) & 0.295 (0.124, 0.701) \\
    p-value & 0.005  & 0.006  & 0.009  & 0.006  \\
    \hline
    \end{tabular}%

         \vspace{2pt} 
  \raggedright\footnotesize
  Abbreviations: SE, standard error; NA, not applicable.\\
\textsuperscript{(a)} The predicted power is monitored using the power formula. For the proposed monitoring procedure, on the day the predicted power exceeded the target power 0.8 for the first time, we conducted the final analysis.
  \label{tab:application}%
\end{table}%

\clearpage

\end{document}